\documentclass[a4paper,10pt]{article}
\usepackage{amsfonts,latexsym}
\usepackage{amsmath}
\usepackage{amscd}
\usepackage{float,amsmath,amssymb,mathrsfs,bm,multirow,graphics}
\usepackage[dvips]{graphicx}

\def\b{\begin{eqnarray}}
\def\e{\end{eqnarray}}
\def\n{\noindent}
\newcommand{\nequation}{\setcounter{equation}{0}}

\newcommand{\C}{{\Bbb C}}

\newcommand{\Z}{{\Bbb Z}}

\newtheorem{theorem}{Theorem}[section]
\newtheorem{proposition}[theorem]{Proposition}

\begin{document}

\begin{center}

{\LARGE\textbf{Inverse Scattering Transform for the
Degasperis-Procesi Equation
\\}} \vspace {10mm} \vspace{1mm} \noindent

{\large \bf Adrian Constantin$^{a,\dag}$}, {\large \bf  Rossen I.
Ivanov$^{b,\ddag}$} and \\ {\large \bf Jonatan Lenells$^{c,\ast}$}

\vskip1cm

\n

\hskip-.3cm
\begin{tabular}{c}
\hskip-1cm $\phantom{R^R} ^{a}${\it Faculty of Mathematics,
University of Vienna, Nordbergstrasse 15,}
\\ {\it 1090 Vienna, Austria} \\
\\
$\phantom{R^R}^{b}${\it School of Mathematical Sciences, Dublin Institute of Technology, Kevin Street}\\ {\it Dublin 8, Ireland} \\
\\
$\phantom{R^R}^{c}${\it Institut f\"ur Angewandte Mathematik, Leibniz Universit\"at Hannover, }\\
{\it Welfengarten 1, 30167 Hannover, Germany}
\\
\\{\it $^\dag$e-mail: adrian.constantin@univie.ac.at}
\\{\it $^\ddag$e-mail: rivanov@dit.ie}
\\{\it $^\ast$e-mail: lenells@ifam.uni-hannover.de}
\\
\hskip-.8cm
\end{tabular}
\vskip1cm
\end{center}
\input epsf

\begin{abstract}
\noindent We develop the Inverse Scattering Transform (IST) method
for the Degasperis-Procesi equation. The spectral problem is an
$\mathfrak{sl}(3)$ Zakharov-Shabat problem with constant boundary
conditions and finite reduction group. The basic aspects of the
IST such as the construction of fundamental analytic solutions,
the formulation of a Riemann-Hilbert problem, and the
implementation of the dressing method are presented.
\end{abstract}

\section{Introduction}\nequation
The Degasperis-Procesi (DP) equation
\begin{equation}\tag{DP}\label{DP}
  u_t - u_{txx} + 3\kappa u_x+4uu_x - 3u_xu_{xx} - uu_{xxx}=0,
\end{equation}
where $\kappa > 0$ is a constant, was first discovered in \cite{D-P} in a search for asymptotically integrable PDEs.
Equation (\ref{DP}) is a bi-Hamiltonian system and admits interesting traveling wave solutions \cite{D-H-H}. It arises as a model equation in the study of two-dimensional water waves
propagating over a flat bed \cite{J, CL, I2}. Given the intricate structure of the full governing equations for water waves, it is natural to seek, in various physical regimes, simpler approximate model equations. In the case of two-dimensional waves propagating mainly in one direction, two parameters appear in the non-dimensionalisation of the governing
equations: the wave-amplitude parameter $\varepsilon$ (indicating
how close the waves are to a flat-surface flow) and the long-wave parameter $\delta$
(measuring the ratio of the approximate wavelength to the average water depth) \cite{J0}.
The relative sizes of these two fundamental parameters determine the different physical
regimes for water waves. The most studied regime is the shallow-water regime (also called the small amplitude
or long-wave regime) for which $\delta \ll 1$ and $\varepsilon \sim \delta^2$. In this parameter range, due to a balance between nonlinearity and dispersion, various integrable systems like the Korteweg-de Vries (KdV) equation arise as approximations to the governing equations \cite{AL}.
The integrability of these equations implies that the powerful method of inverse
scattering can be used to obtain detailed qualitative (and even quantitative) information
about the wave dynamics. In particular, since linear
water wave theory cannot provide an approximation of solitary waves
(see the discussion in \cite{CE2}), an important outcome of studies of the KdV equation was
a deeper understanding of the dynamics of solitary water waves \cite{J0}.

The shallow amplitude regime is, however, not appropriate for the study of large amplitude waves, whose behavior is more nonlinear
than dispersive. To model such waves, which are characterized by a relatively large value of $\varepsilon$, it is
natural to investigate the parameter regime in which $\delta \ll 1$ and $\varepsilon \sim \delta$.
Also in this parameter range is a reduction to a simple wave
equation at leading order possible, but since the dimensionless parameter $\varepsilon$ is larger, the nonlinear effects are stronger than in the shallow water regime. Of particular interest in this regard is that a stronger nonlinearity could allow for the occurence of wave-breaking---a fundamental phenomenon in the theory of water waves that is not captured by the
KdV equation (see the discussion in \cite{CE}). In the shallow-water, moderate amplitude
regime [$\delta \ll 1$, $\varepsilon \sim \delta$] several model equations can be derived as
approximations to the governing equations for water waves. However, among the various equations
that arise in this way, there are only two which admit a bi-Hamiltonian structure (see \cite{CL, I}): the
Camassa-Holm (CH) equation \cite{C-H} and the DP equation (\ref{DP}). For these equations, $u(x,t)$ represents the horizontal velocity of the water at a certain depth at time $t$ and position $x$; the depth ratios for CH and DP are different \cite{J, CL} and are encoded in (\ref{DP}) in the positive non-dimensional parameter $\kappa>0$. In the limit $\kappa \to 0$ the
solitary waves of (\ref{DP}) are peakons and are given explicitly by
$$u_c(x,t)=c\,e^{-|x-ct|},\qquad x,\,t \in {\mathbb R},$$
where $c>0$ denotes the speed of the wave \cite{D-H-H, L-S}. Due to the presence of a peak at the wave crest, these waves have to be understood as weak solutions \cite{L}. This feature is
characteristic for waves of great height (i.e. waves of largest
amplitude that are exact solutions of the governing equations for
water waves, see the discussion in \cite{C, CE2, To}). The
dynamics of the peakon interactions for DP was elucidated in
\cite{M1, M2,D-H-H2}.

In this paper, we will develop an inverse scattering approach for smooth
localized solutions to (\ref{DP}). More precisely, we consider solutions $u(x,t)$ of class $C^1$ in
time and of Schwartz class regularity with respect to the spartial $x$-variable (i.e. the solution is smooth and decays to zero faster than any polynomial as $|x| \to \infty$). Moreover, we will assume that the solution satisfies the following inequality initially (at time $t=0$):
\begin{equation}\label{q}
q = u- u_{xx} + \kappa > 0.
\end{equation}
Note that well-posedness and global existence of solutions
for (\ref{DP}) holds within the class of Schwartz functions if the initial data satisfy (\ref{q}), and in this case the validity of (\ref{q}) is ensured at any later time $t > 0$ (see \cite{ELY, H, LY}).

The paper is organized as follows. In Section 2 we present the Lax pair formulation of (\ref{DP}) and study the symmetry properties of the isospectral problem and of the scattering matrix. Section 3
is devoted to the associated Riemann-Hilbert problem, while the
Zakharov-Shabat dressing method is implemented in Section 4.

\section{Spectral problem}

\subsection{Lax pair}

Equation (\ref{DP}) admits the Lax pair formulation
\cite{D-H-H,HW}
\begin{equation}\label{KK}
\begin{cases}
\varphi_{xxx} - \varphi_x - q\zeta^3 \varphi = 0, \\
\varphi_t - \frac{1}{\zeta^3} \varphi_{xx} + u \varphi_x - u_x
\varphi = 0,
\end{cases}
\end{equation}
where $\zeta \in \C$ is the spectral parameter and $\varphi(x,t)$
is a scalar function. Third-order spectral problems appear as
eigenvalue problems for the Boussinesq, the Sawada-Kotera and the
Kaup-Kuperschmidt equation \cite{Ka2} and they were first
investigated by Kaup \cite{Ka} who established the analyticity
properties of the fundamental solutions, constructed scattering
data and presented an inverse scattering approach in the cases
when only the continuous spectrum is present and when only the
bound state spectrum is present. The third order problem
(\ref{KK}) however is a weighted spectral problem which requires
additional care. For example, a weighted second order spectral
problem arises in the inverse scattering for the CH equation, its
analysis is described in details in \cite{CGI}.

Our aim is to develop the inverse scattering approach for
(\ref{DP}). The Lax representation (\ref{KK}) however is
inconvenient for this purpose. It turns out that the problem
simplifies a lot if the Lax representation is written in the form
of Zakharov-Shabat (ZS) type spectral problem
\cite{ZS1,Sh79,Sh75,ZMNP,FaTa,GVY,VSG04,VSG86,VSG1}. This allows
us to take advantage of the existing inbuilt
symmetries of the considered equation. For this reason we write the above Lax pair in matrix form as
\begin{equation}\label{matrixlax}
\begin{cases}
\phi_{x} = \tilde{L}\phi, \\
\phi_t = \tilde{M} \phi,
\end{cases}
\end{equation}
where
$$\tilde{L} = \begin{pmatrix}  -1 & \zeta  & 0 \\
 0 & 0 & \zeta  \\
 \zeta  q & 0 & 1 \end{pmatrix},Ê
 \qquad
 \tilde{M} = \begin{pmatrix}
  u+u_x + \frac{1}{3 \zeta^3} & -\zeta u-\frac{1}{\zeta^2} & \frac{1}{\zeta } \\
  \frac{q+u_x+u_{xx}}{\zeta } & -\frac{2}{3 \zeta ^3} & \frac{1}{\zeta ^2}-\zeta  u \\
 \frac{-q u \zeta^3 + q_x - u_x + u_{xxx}}{\zeta ^2} & \frac{q + u_x + u_{xx}}{\zeta }
 & -u - u_x + \frac{1}{3 \zeta^3}
   \end{pmatrix}
 $$
and $\phi(x,t)$ is a $SL(3)$ - matrix-valued function, whose
columns, considered as vectors, represent the three linearly
independent solutions of the matrix equation. Let $G(x,t)$ be a
$SL(3)$ matrix and let us take a change of variables $\phi = G
\psi$. It transforms (\ref{matrixlax}) into
$$\begin{cases}
\psi_{x} = L\psi, \\
\psi_t = M \psi,
\end{cases}$$
where
$$L = G^{-1} \tilde{L} G - G^{-1} G_x, \qquad M = G^{-1} \tilde{M} G - G^{-1} G_t.$$
Letting $\omega = e^{2 \pi i/3}$ and
$$G = \frac{1}{\sqrt{3}}
\begin{pmatrix}
 q^{-1/3} & 0 & 0 \\
 0 & 1 & 0 \\
 0 & 0 & q^{1/3} \end{pmatrix}
 \begin{pmatrix}
  1 & 1 & 1 \\
 \omega  & \omega ^2 & 1 \\
 \omega ^2 & \omega  & 1\end{pmatrix},$$
 we find $L = \zeta q^{1/3} J - \tilde{Q}$ where $\tilde{Q}=Q^* \left(1 - \frac{q_x}{3q}\right)$
 \begin{equation}\label{Jqdef}
 J = \begin{pmatrix} \omega & 0 & 0 \\
 0 & \omega^2 & 0 \\
 0 & 0 & 1\end{pmatrix} \quad \hbox{and}\quad
 Q^* = \frac{1}{3}(1-\omega)\begin{pmatrix}  0 & \omega +1 & 1 \\
 1 & 0 & \omega +1 \\
 \omega +1 & 1 & 0\end{pmatrix} .
 \end{equation}

 Let us change the variables according to

\b \label{y} y= x+\int_{-\infty} ^{x} \Big[
\Big(\frac{q(x')}{\kappa}\Big)^{1/3}-1 \Big]\text{d}x', \qquad
\frac{\text{ d} y }{\text{ d} x}
=\Big(\frac{q(x)}{\kappa}\Big)^{1/3}.\e

\n  The $t$-variable can be viewed as an additional parameter
rather than a second independent variable. This will be clear in
the following section and is due to the fact that the
$t$-dependence of the scattering data can be explicitly computed
in relatively simple form. For the sake of simplicity in what
follows we usually omit the $t$-dependence of the variables,
unless this dependence is necessary for the computations. The
spectral problem
\begin{equation}\label{newxpart}
  \psi_x + (\tilde{Q} - \zeta q^{1/3} J ) \psi = 0
\end{equation}

\n can be written in the form

\begin{equation}\label{SP}
\psi_y + (Q- \lambda J ) \psi = 0,
\end{equation} where \b \label{h} \lambda=\zeta \kappa^{1/3}, \quad Q=Q^* h, \quad
h(x)=\Big(\frac{q(x)}{\kappa}\Big)^{-1/3}+\frac{\text{ d} }{\text{
d} x}\Big(\frac{q(x)}{\kappa}\Big)^{-1/3}, \e

\n and $\lim_{x\rightarrow \pm \infty} h(x)=1$.

Suppose that $x=X(y)$. It is possible to recover $q(x)$ from
$h(X(y))$. First, we notice that asymptotically $y\rightarrow x$
when $x\rightarrow - \infty$. Since
$$\int_{-\infty} ^{\infty}\Big[ \Big(\frac{q(x)}{\kappa}\Big)^{1/3}-1 \Big]\text{d}x$$
is an integral of motion \cite{D-H-H}, for $x\rightarrow \infty$ we have that $x$
and $y$ differ only by a constant. Let us introduce for
convenience

\b \label{f} f(y)=\Big(\frac{q(X(y))}{\kappa}\Big)^{1/3}.\e

\n Then $f$, and therefore $q$, can be recovered from $h(y)\equiv
h(X(y))$ by solving the first order differential equation that
follows from (\ref{h}):

\b \label{ODE} \frac{\text{ d}f }{\text{ d} y}+ h(y)f =1, \qquad
f(\pm \infty)=1. \e

\n We obtain the solution in parametric form

\b \label{m-parametric} q(X(y))= \kappa f^3(y), \qquad x\equiv
X(y)=y+ \int_{-\infty}^y\Big(\frac{1}{f(y')}-1\Big) \text{d}y', \e

\n or, with the convolution interpreted in the sense of
distributions \cite{Ho}, \b \label{m} q(x)=\kappa \int_{-\infty}
^{\infty} \delta(x-X(y))f^2(y) \text{d}y, \e  where the function
$X(y)$ is defined in (\ref{m-parametric}) and $f(y)$ is the
solution of (\ref{ODE}). Finally, $h(y)$, which is actually
$h(X(y))$, can be obtained from the scattering data of the
spectral problem (\ref{SP}) which is a ZS-type spectral problem,
however with constant boundary conditions, since $Q(y)=Q^* h(y)$
and $\lim_{y\rightarrow \pm \infty} h(y)=1$. For other ZS
spectral problems for multicomponent systems with constant
boundary conditions we refer to \cite{AG06,GK83,FaTa}.

\subsection{Automorphisms}

The specific form of the potential $Q$ in (\ref{SP}) is due to the
symmetry of the problem under the action of three distinct
automorphisms. In other words, the fact that $Q$ is determined by
a single real (scalar) function, rather than 6 complex functions
(which is the case for an arbitrary $sl(3)$ potential) is a
consequence of its invariance under one $\Z_3$ automorphism and
two $\Z_2$ automorphisms. The automorphisms lead to the reduction
of the independent components of the potential and their action
extends to the spectrum and the eigenfunctions. They form a group,
known as \emph{a reduction group}
\cite{AV,GKV,GVY,GGK05a,GG1,GG2,G1,G2,G3,G4}.

\subsubsection{$\Z_3$ automorphism} The spectral problem (\ref{SP})
has a manifest $\Z_3$ symmetry:

\b C L(\omega \lambda)C^{-1} = L(\lambda), \label{Z3}\e

\n where
$$C =  \begin{pmatrix}
0 & 0 & 1 \\
 1 & 0 & 0 \\
 0 & 1 & 0\end{pmatrix}. $$
Indeed, one can check that
$$CJC^{-1} = \frac{1}{\omega} J, \qquad CQC^{-1} = Q$$
\n from where (\ref{Z3}) follows immediately. Furthermore, one can verify that
\begin{equation}\label{Msymmetry}
  C M(\omega \lambda)C^{-1} = M(\lambda),
\end{equation}
so that the $\Z_3$ automorphism (\ref{Z3}) is an automorphism of
the graded Kac-Moody algebra where $L(\lambda)$ and $M(\lambda)$
take their values \cite{Ger}. This holds for all
automorphisms of the spectral problem.

\subsubsection{$\mathbb{Z}_2$ automorphisms}

The spectral problem possess two additional $\mathbb{Z}_2$
automorphisms, one of which reflects the reality of $h(y)$. The
first one is

\b \Gamma \overline{L(\overline{\omega \lambda})}\Gamma^{-1} =
L(\lambda), \label{Z2}\e

\noindent where

$$\Gamma =  \begin{pmatrix}
0 & 0 & 1 \\
0 & 1 & 0 \\
1 & 0 & 0\end{pmatrix}.
$$

\noindent
The second $\Z_2$ automorphism is

\b Z L^{\dag}(-\bar{\lambda})Z^{-1} = -L(\lambda), \label{Z2-2}\e

\noindent where

$$ Z =  \begin{pmatrix}
0 & \omega & 0 \\
\omega^2 & 0 & 0 \\
0 & 0 & 1\end{pmatrix},
$$
and the dagger stands for a matrix Hermitian conjugation.

One can easily check that all $\Z_2$ and $\Z_3$ automorphisms are
compatible, in the sense that the order of their application to
the potential does not matter.

\section{Construction of Fundamental Analytic Solutions}

\subsection{Asymptotic behavior}
Let us define the asymptotic values
$$L_\infty (\lambda) = \lim_{y \to \pm \infty} L(y,\lambda), \qquad M_\infty (\lambda)= \lim_{y \to \pm \infty} M(y,\lambda).$$
Then
$$L_\infty = \lambda  J - Q^*$$
where $Q^* $ is defined in (\ref{Jqdef}). We also find that \b
M_\infty (\lambda)= \frac{\kappa}{3 \lambda^3}
\begin{pmatrix}
 3 \omega ^2 \lambda ^2 &
   \omega ^2 \left( \omega -1
   \right) \lambda -\omega &   \omega ^2 (1
   -\omega ) \lambda -\omega ^2 \\
  (1- \omega) \lambda -\omega ^2   & 3 \omega
   \lambda ^2    &  (\omega  -1 ) \lambda - \omega  \\
 \omega( \omega -1) \lambda -\omega & \omega \left(1- \omega \right) \lambda - \omega ^2&
  3\lambda ^2
   \end{pmatrix} \label{M_infty}\e  and that
$$[L_\infty , M_\infty ] = 0.$$
Since $L_\infty$ and $M_\infty$ commute, they can be
simultaneously diagonalized. Let $U(\lambda)$ be a $SL(3)$ matrix
such that
$$L_\infty (\lambda)  = U (\lambda)  \Lambda (\lambda) U^{-1}(\lambda) ,
\qquad M_\infty(\lambda)  = U(\lambda)  A(\lambda) U^{-1}(\lambda)
,$$
with
$$
\Lambda(\lambda)  = \text{diag}(\Lambda_1(\lambda) ,
\Lambda_2(\lambda) , \Lambda_3(\lambda) ),\qquad
A(\lambda) =\text{diag}(A_1(\lambda) , A_2(\lambda) , A_3(\lambda) ),$$
where $\Lambda_1, \Lambda_2, \Lambda_3$ and $A_1, A_2, A_3$ are the
eigenvalues of $L_\infty$ and $M_\infty$ respectively.

The eigenvalues of $L_{\infty}(\lambda)=\lambda J-Q^*$  are the
solutions $\Lambda(\lambda)$ of the characteristic equation
$$\Lambda^3-\Lambda -\lambda^3 = 0.$$

\n Introducing a new spectral parameter $k$ such that \b
\lambda(k)=3^{-1/2}k\Big(1+\frac{1}{k^6}\Big)^{1/3}=3^{-1/2}k\left(1+\frac{1}{3k^6}+\ldots\right)
\label{lambda(k)}\e we have \b
\lambda^3=3^{-3/2}\Big(k^3+\frac{1}{k^3} \Big)\e and the following
expression for the eigenvalues of $L_{\infty}$: \b \label{Lambdas}
\Lambda_j (k)=3^{-1/2}\Big(\omega^j k+\frac{1}{\omega^j k}\Big).
\e

\n Furthermore, $\lambda(k)$ has the property $\lambda(\omega
k)=\omega \lambda(k)$ and also \b \lambda(k)\rightarrow 3^{-1/2}k
\qquad \text{when} \qquad |k| \rightarrow \infty. \e

The characteristic polynomial of the matrix $3\kappa^{-1}\lambda
^3 M_{\infty}$,  cf. (\ref{M_infty}), is \b P(w)=w^3-3w-27
\lambda^{6} +2, \label{char poly M} \e

\noindent with roots
$$w_{j}(k)=\omega^{j}k^2+\omega^{-j}k^{-2},$$
where $\lambda(k)$ is given in (\ref{lambda(k)}). Thus the
eigenvalues of $M_{\infty}$ are  \b A_{j}(k)=\frac{\kappa
w_{j}(k)}{3 \lambda^3(k)} \label{mu=s/lamda}. \e

\noindent It remains to determine the ordering of the eigenvalues
$A_j$ that is consistent with the ordering of $\Lambda_j$, the
eigenvalues of $L_{\infty}$. To this end we consider the
asymptotic expressions when $k\rightarrow \infty$. Then

\b L_{\infty}\rightarrow \frac{k J}{\sqrt{3}}, \qquad
M_{\infty}\rightarrow \frac{\sqrt{3}\kappa}{k}{J ^2 }, \nonumber\e

\noindent or

\b \Lambda_j \rightarrow \frac{k\omega^j}{\sqrt{3}}, \qquad
A_{j}\rightarrow \frac{\sqrt{3}\kappa}{k}\omega ^{2j}.\e

\noindent Thus, from (\ref{mu=s/lamda}) we have

\b A_j(k)&=&\sqrt{3} \kappa \frac{(\omega^j k)^{2}+(\omega^j
k)^{-2}}{k^3+k^{-3}};  \label{mu_n_k} \e

\subsection{Scattering matrix}

Consider the modified Lax pair
\begin{equation}\label{laxpair}
  \begin{cases}
    \psi_y = L\psi, \qquad L=\lambda J - Q(y) \\
    \psi_t = M\psi - \psi M_\infty(\lambda).
  \end{cases}
\end{equation}
The compatibility condition holds for any choice of matrix replacing $M_\infty(\lambda)$, i.e. the modified Lax pair gives
rise to the same equation as the original Lax pair. For
convenience we use the spectral parameter $k$. Let
$\psi^\pm(y,t,k)$ be the solutions of (\ref{laxpair}) such that
$$\lim_{y \to \pm \infty}\psi^\pm(y,t, k) = U(k) e^{\Lambda(k) y} U^{-1}(k). $$

Note that due to the modified second equation in (\ref{laxpair})
the asymptotic values do not depend on $t$. The two solutions
$\psi^+(y,t,k)$ and $\psi^-(y,t,k)$ are not linearly independent,
i.e. they are related by a linear transformation when $k$ is a point of the spectrum. Thus, the
expression $\hat{\psi}^+(y,t, k) \psi^-(y,t, k)$ depends on $t$
and $k$, but not on $y$. (From now on we write $\hat{B}$ for the
inverse of a matrix $B$). We define the scattering matrix $T(t,k)$
by
\begin{equation}\label{Tdef}
  T(t,k) = \hat{U}(k) \hat{\psi}^+(y,t, k) \psi^-(y,t, k) U(k),
\end{equation}

The explicit form of the $t$-dependence of $T$ is quite simple:

\begin{proposition}\label{T-t-evol} The time-evolution of the
scattering matrix is given by \b \label{T(t)} T(t,k)=e^{A(k)
t}T(0,k)e^{-A(k) t}. \e
\end{proposition}

{\it Proof.} From (\ref{Tdef}) it follows that
\begin{equation}\label{psipluspsiminus}
  \psi^+ U T = \psi^- U.
\end{equation}
Differentiating both sides with respect to $t$, we obtain
$$\psi^{+}_t U T + \psi^{+} U T_t = \psi^-_{t} U.$$
Replacing $\psi^{\pm }_t$ from the $t$-part of the Lax pair
(\ref{laxpair}), we find
$$(M\psi^+ - \psi^+ M_\infty) U T + \psi^{+} U T_t = (M\psi^- - \psi^- M_{\infty}) U.$$
In view of (\ref{psipluspsiminus}) this becomes
$$-\psi^+ M_\infty U T + \psi^{+} U T_t = -\psi^+ U T\hat{U} M_\infty U.$$
We conclude that $T(t,k)$ evolves according to $T_t = -[T, \hat{U}
M_\infty U],$ or
$$T_t =  [ A,T].$$
Therefore the time-evolution of the scattering matrix is given by
(\ref{T(t)}).$\hfill\Box$

\subsection{Fundamental analytic solutions}

An important role in the theory of inverse scattering play the so
called \emph{fundamental analytic solutions} (FAS) of the spectral
problem. We will explain the construction of FAS for the system
(\ref{SP}) or rather for the related spectral problem \b \label{xi}
\xi_y+ \tilde{Q}\xi + [\xi, \Lambda(k)]=0, \qquad
\tilde{Q}=\hat{U}(Q-Q^*)U, \label{RelatedSP} \e  in the domains
$\Omega_{\nu}$, $\nu=1, \ldots, 6$ occupying the space
in the complex $k$-plane separated by the rays $\{l_{\nu}$:
$\arg(k)=(\nu-1) \frac{\pi}{3}\}$ with $\nu=1,\,...,\,6$ (see Figure 1). It is easy to notice that the
components of $\tilde{Q}(y)$ are functions in the Schwartz class.

\begin{figure}
\centering
\includegraphics[width=70mm]{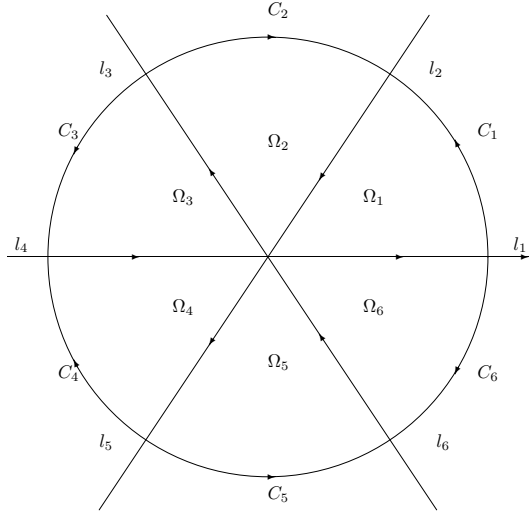}
\caption{Domains of analyticity and integration contours.}
\label{contour1}
\end{figure}

Let us introduce an ordering for each sector $\Omega_{\nu}$ as
follows. We will say that

\b r \mathop{<}\limits_{\nu} s \qquad \text {iff} \qquad \Re[
\Lambda_r(k)]< \Re[ \Lambda_s(k)]\qquad \text{ when} \qquad k\in
\Omega_{\nu}. \e

\n One can verify that
$1\mathop{<}\limits_{1}2\mathop{<}\limits_{1}3$ since for $ k\in
\Omega_1$, $\Re[ \Lambda_1(k)]< \Re[ \Lambda_2(k)]<\Re[
\Lambda_3(k)]$. Similarly,
$1\mathop{<}\limits_{2}3\mathop{<}\limits_{2}2$, etc. Then we can
prove the following result:

\begin{proposition}\label{prop1} The solution $\xi^{\nu}(y, k)$ of
the following system of integral equations

\b \label{xiIntE1}
\xi^{\nu}_{rs}(y,k)\!\!\!&=&\!\!\!\delta_{rs}-\int_{-\infty}^y\!
e^{-[\Lambda_r(k)-\Lambda_s(k)](y-y')}[\tilde{Q}(y',k)\xi^{\nu}(y',k)]_{rs}\text{d}
y', \quad r\mathop{\geq}\limits_{\nu}s;\phantom{****}
\\\label{xiIntE2}
\xi^{\nu}_{rs}(y,k)&=&-\int_{\infty}^y
e^{-[\Lambda_r(k)-\Lambda_s(k)](y-y')}[\tilde{Q}(y',k)\xi^{\nu}(y',k)]_{rs}\text{d}
y', \quad r\mathop{<}\limits_{\nu}s. \e

\noindent
is also a solution of spectral problem (\ref{xi}) and is analytic
for $k\in \Omega_{\nu}$.
\end{proposition}

{\it Proof.} The first part of the statement follows by direct
computation. The analyticity follows from the fact that the real
part of $\Lambda_r(k)-\Lambda_s(k)$ is nonnegative when $k\in
\Omega_{\nu}$ and $r\mathop{\geq}\limits_{\nu}s$ etc.$\hfill\Box$\bigskip

By direct computation we prove the following useful result.\bigskip

\begin{proposition}\label{lema} The solution $\xi(y, k)$ of
(\ref{xi}) is related to a solution $\psi(y,k)$ of (\ref{SP}) via
$$\xi(y,k)=\hat{U}(k)\psi(y,k) U(k) B(k) e^{-\Lambda(k) y}$$
where $B(k)$ is some nondegenerate matrix. \end{proposition}

Using the two propositions we compute the asymptotics
\b \lim_{y\rightarrow \infty} [e^{-\Lambda(k)
y}\xi^{\nu}(y,k)e^{\Lambda(k) y}]_{rs}=0, \qquad
r\mathop{<}\limits_{\nu}s, \qquad k\in \Omega_{\nu}. \nonumber \e
Therefore \b \lim_{y\rightarrow \infty} e^{-\Lambda(k)
y}\xi^{\nu}(y,k)e^{\Lambda(k) y}=T^-_{\nu}(k)D^+_{\nu}(k), \qquad
k\in \Omega_{\nu}, \label{asympt inf} \e

\n where $T^-_{\nu}(k)$ is a lower -triangular matrix with units
on the diagonal with respect to the $\nu$-ordering and
$D^+_{\nu}(k)$ is a diagonal matrix. Similarly,
\b \lim_{y\rightarrow -\infty} e^{-\Lambda(k)
y}\xi^{\nu}(y,k)e^{\Lambda(k) y}=S^+_{\nu}(k), \qquad k\in
\Omega_{\nu}, \label{asympt -inf} \e

\n where $S^+_{\nu}(k)$ is an upper-triangular matrix with units
on the diagonal with respect to the $\nu$-ordering.

A more-subtle result, the proof of which follows exactly the lines of the proof of Theorem
3.3. from \cite{GY}, gives the asymptotics for $k\in l_{\nu}$:

\begin{proposition}\label{lema3.3GY}
When $k\in l_{\nu}$, both $[S^+_{\nu}(k)]_{rs}=0$ and
$[T^-_{\nu}(k)]_{rs}=0$ if
$\Re(\Lambda_r(k))\ne\Re(\Lambda_s(k))$.
\end{proposition}

In other words, on the line $k\in l_{\nu}$ nonzero could be only
the entries that have the numbers $r_0$, $s_0$ of the two
eigenvalues that define $l_{\nu}$ as
$\Re(\Lambda_{r_0}(k))=\Re(\Lambda_{s_0}(k))$. For example, when
$\nu=2$, $r_0=2$, $s_0=3$. Notice that if in $\Omega_{\nu-1}$
$r_0\mathop{<}\limits_{\nu-1}s_0$, in the neighboring
$\Omega_{\nu}$ ($l_{\nu}$ separates $\Omega_{\nu-1}$ and
$\Omega_{\nu}$) the ordering changes such that
$r_0\mathop{>}\limits_{\nu}s_0$. Therefore, if the matrix
$S^+_{\nu-1}(k)$ is upper triangular with respect to the $\nu-1$
ordering, it is lower-triangular with respect to the
$\nu$-ordering and we will denote it as $S^-_{\nu}(k)$, etc.

The asymptotics on the ray $k\in l_{\nu}$ therefore can be written
in the form: \b  \lim_{y\rightarrow \infty} e^{-\Lambda(k)
y}\xi^{\nu}(y,k)e^{\Lambda(k)
y}&=&T^-_{\nu}(k)D^+_{\nu}(k),  \label{asymptinf nu} \\
\lim_{y\rightarrow -\infty} e^{-\Lambda(k)
y}\xi^{\nu}(y,k)e^{\Lambda(k) y}&=&S^+_{\nu}(k),  \label{asympt
-inf nu} \\
\lim_{y\rightarrow \infty} e^{-\Lambda(k)
y}\xi^{\nu-1}(y,k)e^{\Lambda(k)
y}&=&T^+_{\nu}(k)D^-_{\nu}(k),  \label{asymptinf nu-1} \\
\lim_{y\rightarrow -\infty} e^{-\Lambda(k)
y}\xi^{\nu-1}(y,k)e^{\Lambda(k) y}&=&S^-_{\nu}(k),
\label{asympt-inf nu-1} \e

\n where the matrices $S^+_{\nu}$, $T^+_{\nu}$ (resp. $S^-_{\nu}$,
$T^-_{\nu}$ ) are upper-triangular (resp. lower triangular) with
units on the diagonal (with respect to the $\nu$-ordering) and the
matrices $D^{\pm}$ are diagonal. They provide the Gauss
decomposition of the scattering matrix with respect to the
$\nu$-ordering \cite{Ger,GY}, i.e.

\begin{equation}\label{gaussdecomposition}
  T_{\nu} (k) = T^-_{\nu}(k)D^+_{\nu}(k)\hat{S}^+_{\nu}(k) = T^+_{\nu}(k) D^-_{\nu}(k)
  \hat{S}^-_{\nu}(k),\qquad k\in l_{\nu}
\end{equation}
where $T^-, S^-$ are lower-triangular matrices with units along
the diagonal, $T^+, S^+$ are upper-triangular matrices with units
along the diagonal, and $D^\pm$ are diagonal matrices. From the
definition (\ref{Tdef}) of $T$ and (\ref{gaussdecomposition}) it
follows that on the lines $k\in l_{\nu}$, \b \psi^+(y,k) U(k)
T^-_{\nu}(k) D^+_{\nu}(k) &=& \psi^-(y,k) U(k) S^+_{\nu}(k),\\
\psi^+(y,k) U(k) T^+_{\nu}(k) D^-_{\nu}(k) &=& \psi^-(y,k) U(k)
S^-_{\nu}(k). \label{psi pm and T factors 2}\e

\n Now, using the results from Proposition \ref{lema} and
(\ref{asymptinf nu})-- (\ref{psi pm and T factors 2}) we can
relate the eigenfunctions $ \xi^{\nu}$ with $\psi^{\pm}$ for $k\in
l_{\nu}$:

\b  \xi^{\nu}(y, k)&=&\hat{U}(k)\psi^-(y,k)U(k)
S^+_{\nu}(k)e^{-\Lambda(k) y}\nonumber \\ &=&
\hat{U}(k)\psi^+(y,k) U(k) T^-_{\nu}(k)D^+_{\nu}(k)e^{-\Lambda(k)
y} \label{xi_nu} \e

\b   \xi^{\nu-1}(y, k)&=&\hat{U}(k)\psi^-(y,k)U(k)
S^-_{\nu}(k)e^{-\Lambda(k)
y}\nonumber \\
&=&\hat{U}(k)\psi^+(y,k) U
(k)T^+_{\nu}(k)D^-_{\nu}(k)e^{-\Lambda(k) y}.
 \label{xi_nu-1}\e

The symmetries  (\ref{Z3}), (\ref{Z2}) and (\ref{Z2-2}) impose
constraints to the eigenfunctions, the scattering matrix and its
factors e.g. \b C\xi^{\nu+2}(y, \omega k)\hat{C}&=&\xi^{\nu}(y,
k), \label{sym1}\e
\b \Gamma \bar{\xi}^{5-\nu}\left(y, \bar{\omega}
\bar{k}\right)\hat{\Gamma}&=& \xi^{\nu}(y,
k), \label{sym11}\e
\b Z [\xi^{4-\nu}\left(y,
-\bar{k}\right)]^{\dag}\hat{Z}&=&\hat{\xi}^{\nu}(y,
k), \label{Z2-2b}\e etc.  where the $\nu$-indices should be taken modulo 6.

Thus, independent is only the data on one of the lines, say $l_0$,
and all the rest can be recovered from (\ref{sym1}) -- (\ref{Z2-2b}).

\section{Riemann-Hilbert problem}
\nequation

Any two solutions $\xi^{\nu}(y,k;t)$ and $\xi^{\nu-1}(y,k;t)$ are
related on $k\in l_{\nu}$ due to (\ref{xi_nu}) -- (\ref{xi_nu-1}):

\b \xi^{\nu}(y,k;t)&=&\xi^{\nu-1}(y,k;t) G^{\nu}(y,k;t), \label{RHP1}\\
G^{\nu}(y,k;t)&=&e^{\Lambda(k) y + A
(k)t}\hat{S}^{-}_{\nu}(k)S^+_{\nu}(k)e^{-\Lambda(k) y- A
(k)t}\qquad k\in l_{\nu}, \label{RHP2}\\
\lim_{k\rightarrow  \infty} \xi^{\nu}(y,k;t)&=&1\!\!\text{I}.
\label{RHP3} \e

We can show that the relations (\ref{RHP1}) -- (\ref{RHP3})
constitute a Riemann-Hilbert problem (RHP) for the matrix-valued
functions $\xi^{\nu}(y,k;t)$, each being analytic for $k\in
\Omega_{\nu}$. In other words, the solution of the original
Inverse Scattering Problem reduces to a RHP

\begin{proposition}\label{propRHP} The RHP (\ref{RHP1}) --
(\ref{RHP3}) has a unique solution $\xi^{\nu}(y,k;t)$, with an analytic continuation for $k\in \Omega_{\nu}$.

\end{proposition}

For simplicity we consider the case where $\xi^{\nu}(y,k;t)$ do
not have any singularities in the $k$-plane. Suppressing $y$ and
$t$-dependence for convenience where possible, we can write an
analytic continuation for $\xi^{\eta}(y,k)$ ($k \in
\Omega_{\eta}$) as follows:

\b \xi^{\eta}(y,k)=\sum_{\nu=1}^{6}\frac{(-1)^{\nu+1}}{2\pi i}
\oint_{\partial \Omega_{\nu}} \frac{\xi^{\nu}(y,k')\text{d}
k'}{k'-k},\e

\noindent where the orientations of the contours $\partial
\Omega_{\nu}$ are shown on Figure 1 and $C_{\nu}$ belong to the
infinite circle. Due to (\ref{RHP3}) we have

\b \xi^{\eta}(y,k)&=& \frac{1}{2\pi i}
\sum_{\nu=1}^{6}\left(\int_{l_{\nu}} \frac{\xi^{\nu}(y,k')\text{d}
k'}{k' - k}-\int_{l_{\nu+1}} \frac{\xi^{\nu}(y,k')\text{d} k'}{k'
-k} + (-1)^{\nu+1}\int_{C_{\nu}}\frac{\!\!1\!\!\text{I}\text{d}
k'}{k'-k}\right) \nonumber \\
\!\!&=&\!\!1\!\!\text{I}+\frac{1}{2\pi i}\sum_{\nu=1}^{6}
\left(\int_{l_{\nu}} \frac{\xi^{\nu}(y,k')-\xi^{\nu-1}(y,k')}{k' -
k}\text{d}k'\right), \label{RHP4} \e

\noindent where the orientation of $l_{\nu}$ is always from $0$ to
$\infty$. From (\ref{RHP1}) it follows that

\b \xi^{\eta}(y,k)\!\!&=&\!\!1\!\!\text{I}+\frac{1}{2\pi i}
\left(\sum_{\nu=1}^{6}\int_{l_{\nu}}
\frac{\xi^{\nu-1}(y,k')[G^{\eta}(y,k')-\!\!1\!\!\text{I}]\text{d}
k'}{k' - k}\right), \qquad k\in \Omega_{\eta}. \label{RHP6} \e

\noindent If $k$ approaches $l_{\eta}$ from the left and right
domains correspondingly, with the Sohotski-Plemelj-type formulae
we obtain that

\b
\xi^{\eta}(k)\!\!&=&\!\!1\!\!\text{I}+\frac{1}{2}\xi^{\eta-1}(k)[G^{\eta}(k)-\!\!1\!\!\text{I}]
+\frac{1}{2\pi i} \left(\sum_{\nu=1}^{6}P.V.\int_{l_{\nu}}
\frac{\xi^{\nu-1}(k')[G^{\nu}(k')-\!\!1\!\!\text{I}]\text{d}
k'}{k' -
k}\right), \nonumber \\
\xi^{\eta-1}(k)\!\!&=&\!\!1\!\!\text{I}-\frac{1}{2}\xi^{\eta-1}(k)[G^{\eta}(k)-\!\!1\!\!\text{I}]
+\frac{1}{2\pi i} \left(\sum_{\eta=1}^{6}P.V.\int_{l_{\nu}}
\frac{\xi^{\nu-1}(k')[G^{\nu}(k')-\!\!1\!\!\text{I}]\text{d}
k'}{k' -
k}\right), \nonumber \\
\hfill  k &\in& l_{\eta},\qquad  \eta=1,2\ldots,6\label{RHP7} \e

\noindent From (\ref{RHP7}) and (\ref{RHP1}) we finally obtain a
system of integral equations for $\xi^{\eta-1}(y,k)$:

\b
\frac{1}{2}\xi^{\eta-1}(y,k)[G^{\nu}(y,k)+\!\!1\!\!\text{I}]\!\!&=&\!\!1\!\!\text{I}
+\frac{1}{2\pi i} \left(\sum_{\nu=1}^{6}P.V.\int_{l_{\nu}}
\frac{\xi^{\nu-1}(k')[G^{\nu}(y,k')-\!\!1\!\!\text{I}]\text{d}
k'}{k' - k}\right), \nonumber \\   k &\in& l_{\eta}, \qquad
\eta=1,2\ldots,6. \label{RHP8} \e

The solutions of (\ref{RHP8}) provide $\xi^{\eta-1}(y,k)$ when
$k\in l_{\eta}$ and then the analytic continuation is given by
(\ref{RHP6}). Apparently the scattering data are provided by
$G^{\nu}(y,k;t)$ given in (\ref{RHP2}), i.e. by $S^{-}_{\nu}(k) $
and $S^+_{\nu}(k)$, $k\in l_{\nu}$. Using the $\mathbb{Z}_2$ and
$\mathbb{Z}_3$ automorphisms (\ref{Z3}) -- (\ref{Z2-2}) i.e.
(\ref{sym1}) -- (\ref{Z2-2b}) we can restrict the scattering data
to $S^{-}_{1}(k) $ and $S^+_{1}(k)$, $k\in l_{1}$. However,
according to Proposition \ref{lema3.3GY}, $S^+_{1}(k)$ has only
one nontrivial component $[S^+_{1}(k)]_{12}$, which is exactly the
only nontrivial component of $S^-_{1}(k)$. Again, the
automorphisms (\ref{sym1}) -- (\ref{Z2-2b}) relate $S^-_{2}(k)$ to
$S^-_{1}(k)$ which restricts the minimal set of scattering data to
only one function, say $[S^+_{1}(k)]_{12}$. This function is the
analogue of the {\it reflection coefficient} used for example in
the Inverse Scattering for the KdV equation \cite{ZMNP}.

The potential $Q(y,t)$ of the scattering problem can be recovered
from the following result:

\begin{proposition}\label{propRHP-potential} If $\xi^{\nu}(y,k;t)$
are the solutions of the RHP (\ref{RHP1}) -- (\ref{RHP3}) then \b
\chi^{\nu}(y,k;t)=U(k)\xi^{\nu}(y,k;t)e^{\Lambda(k) y}
\label{chi-nu}\e satisfy (\ref{SP}) with

\b Q(y,t)=-\lim_{k\rightarrow \infty}\lambda(k)\left(
\chi^{\nu}(y,k;t)J\hat{\chi}^{\nu}(y,k;t)-J\right). \label{Q
recovered}\e

\end{proposition}

{\it Proof.} With arguments similar to those given in
\cite{Sh75,Sh79,Ger} one can prove that \b
\tilde{Q}(y,t)=-\lim_{k\rightarrow  \infty}\lambda(k)\left(
\xi^{\nu}(y,k;t)J\hat{\xi}^{\nu}(y,k;t)-J\right). \label{tilde
Q}\e

\noindent From (\ref{tilde Q}) and the definition of $\tilde{Q}$
in (\ref{RelatedSP}) it follows that

\b Q(y,t)=Q^*-\lim_{k\rightarrow \infty}\lambda(k)\left(
\chi^{\nu}(y,k;t)J\hat{\chi}^{\nu}(y,k;t)-U(k)J\hat{U}(k)\right).
\label{Q 1}\e

\noindent Next, we notice that \b \lim_{k\rightarrow
\infty}\left( \Lambda(k)- \lambda(k)J \right)=0, \nonumber \e

\noindent giving \b \lim_{k\rightarrow  \infty}\left(
U(k)\Lambda(k)\hat{U}(k)- \lambda(k)U(k)J \hat{U}(k)\right)=0,
\nonumber \e

\b \lim_{k\rightarrow \infty}\left( L_{\infty}- \lambda(k)U(k)J
\hat{U}(k)\right)=0, \nonumber \e or  \b \lim_{k\rightarrow
\infty}\left( \lambda (k) J-Q^*- \lambda(k)U(k)J
\hat{U}(k)\right)=0, \nonumber \e i.e.

\b Q^*=\lim_{k\rightarrow \infty}\lambda (k)\left( J- U(k)J
\hat{U}(k)\right). \label{Q star} \e

\noindent Clearly (\ref{Q recovered}) follows from (\ref{Q 1}) and
(\ref{Q star}).$\hfill\Box$\bigskip

{\bf Corollary} {\it From (\ref{Q recovered}) and the fact that
$\text{tr}[(Q^*)^2]=2$, one can find \b
h(y,t)=\frac{1}{2}\text{tr}(Q Q^*) \e and then $f(y,t)$ can be
computed from (\ref{ODE}) and $q(x,t)$ from (\ref{q}). Finally,
$u(x,t)$ can be obtained from (\ref{m-parametric}) and (\ref{m}).}

\section{Zakharov-Shabat dressing method}
\nequation

The Zakharov-Shabat dressing method \cite{Sh75,Sh79,ZaSh1,ZS1}
allows the explicit construction of a
solution with singularities $\chi^{\nu}(y,k;t)$, starting from a
given regular solution of the RHP, say $\chi^{\nu}_0(y,k;t)$:

\b \chi^{\nu}(y,k;t)=g(y,k;t)\chi^{\nu}_0(y,k;t)
\label{dressSolution}.\e
Note that in our case the analyticity regions for $\chi^{\nu}$ do not coincide with those for $\xi^{\nu}$ due to the nontrivial $U(k)$-factors in
(\ref{chi-nu}).

The dressing factor $g$ is analytic in the entire complex plane,
with the exception of the points of the discrete spectrum. We make
the following assumptions in our construction of a dressing
factor.  First, we allow only simple poles of $g$ and $\hat{g}$.
For simplicity, instead of the $k$-dependence we are going to
revert to the $\lambda$ - dependence, having in mind the
previously defined relation $k(\lambda)$. The automorphisms act on
$g(\lambda)$ as on a group element:

\b C g(y,\omega \lambda)\hat{C}=g(y, \lambda), \label{g sym C} \\
\Gamma \bar{g}\left(y, \bar{\omega}\bar{\lambda}
\right)\hat{\Gamma}=g(y, \lambda), \label{sym111} \\
Z g^{\dag}(y,- \bar{\lambda})\hat{Z}=\hat{g}(y, \lambda),
\label{gsym_Z} \e

From these symmetries it follows that if $g$ or $\hat{g}$ have a
pole at, say, $\lambda_0$, then they have also poles at
$-\lambda_0$, $\pm \omega \lambda_0$, $\pm \omega^2 \lambda_0$,
$\pm \bar{\lambda}_{0}$, $\pm \omega \bar{\lambda}_{0}$, $\pm
\omega^2 \bar{\lambda}_{0}$. It also follows that $\det(g)=1$. Our next assumption will be that
$\lambda_0$ can be chosen real, so that the following choice of
$g$ is possible

\b
g(y,\lambda;t)=\!\!1\!\!\text{I}+\sum_{j=1}^{3}\frac{\alpha_j(y,t)}{\lambda-\lambda_{j}},\label{gfactor}\\
\hat{g}(y,\lambda;t)=\!\!1\!\!\text{I}-\sum_{j=1}^{3}\frac{Z\alpha_j^{\dag}(y,t)\hat{Z}}{\lambda-\mu_{j}},
\label{g-hat-factor} \e

\n for some residues $\alpha_j(y,t)$ where

\b \lambda_j=\omega^{j+1}\lambda_0,\qquad \mu_j=-\bar{\lambda}_j \label{lambda-0}.\e
The property (\ref{RHP3}) of the dressed solution is preserved. Following the ZS construction from
\cite{ZS1} (see also \cite{Ger,JC09}), we represent the residues in the form \b \alpha_p=\sum_{j=1}^{3}|n_j\rangle \hat{R}_{jp}\langle m_p| \label{alpha}\\
R_{jp}=\frac{\langle m_j|n_p\rangle}{\lambda_j - \mu_p}
\label{matrixR}\e

\n where $|n_j \rangle $ is a vector-column, and  $\langle m_{j}|$
is a vector-row. The last is defined as \b \langle
m_{j}|&=&\langle m_{0j}|\hat{\chi}^{(j)}(y,\lambda_j).
 \e

\n where $\langle m_{0j}|$ is a constant vector and
$\hat{\chi}^{(j)}$ is the solution of the adjoint problem,
analytic at $\lambda_j$. We define \b |n_j \rangle \equiv
Z\overline{|m_j \rangle}=\chi^{(\tilde{j})}(\mu_j)|n_{0j}\rangle,
\qquad \text{where} \qquad |m_j \rangle = (\langle m_j|)^T \e

\n and $\chi^{(\tilde{j})}(\mu_j)=Z
 [\hat{\chi}^{(j)}(\lambda_j)]^{\dag}\hat{Z}$, i.e. the constant vectors
 are related as follows:

 \b |n_{0j} \rangle =
Z\overline{|m_{0j} \rangle}. \label{n0-m0} \e

\n $\chi^{(\tilde{j})}$ is an eigenfunction, analytic at $\mu_j$,
since it is obtained by the automorphism (\ref{Z2-2}) from
$\hat{\chi}^{(j)}$, analytic at $\lambda_j$ (and $\lambda_j$ and
$\mu_j$ are related by the same automorphism). With these
definitions one can easily check that the matrix $R$
(\ref{matrixR}) is Hermitian: $R=R^{\dag}$.

The residues of $\hat{g}$ in (\ref{g-hat-factor}) can be computed
and represented in the form \b Z\alpha_p^{\dag}(y,t)\hat{Z}
=\sum_{j=1}^{3}|n_p\rangle \hat{R}_{pj}\langle m_j|
\label{alpha-tilde}\e

One can now verify that $g\hat{g}=\!\!1\!\!\text{I}$. Indeed, this
is satisfied iff at any singular point, say $\lambda=\mu_p$ the
corresponding residues satisfy

\b
\left(\!\!1\!\!\text{I}+\sum_{j=1}^{3}\frac{\alpha_j}{\mu_p-\lambda_j}\right)(Z\alpha_p^{\dag}\hat{Z})&=&0,
\nonumber \\
\alpha_p\left(\!\!1\!\!\text{I}-\sum_{j=1}^{3}\frac{(Z\alpha_j^{\dag}\hat{Z})}{\lambda_p-\mu_j}\right)&=&0,
 \label{residues-identity}\e

\n etc. The identities (\ref{residues-identity}) can be verified
by (\ref{alpha}) -- (\ref{alpha-tilde}). Our construction for $g$
(\ref{gfactor}) -- (\ref{g-hat-factor}) satisfies also the
automorphism (\ref{gsym_Z}). The automorphism (\ref{g sym C})
gives an additional relation between the constant vectors
$|n_{0j}\rangle$:

\b |n_{0,j+1}\rangle  =C|n_{0j}\rangle.  \label{n0-C}\e

\n From (\ref{sym111}) we have the further restriction

\b \Gamma \overline{|n_{01}\rangle} = |n_{02}\rangle, \qquad
\Gamma \overline{|n_{03}\rangle} = |n_{03}\rangle.
\label{n0-Gamma}\e

The relations (\ref{n0-C}) show that only one of the vectors, say
$|n_{01}\rangle$, determines the others (and also $\langle m_{0j}|$
due to (\ref{n0-m0})). The components $n_{01;j}$ ($j=1,2,3$) of
the vector $|n_{0,1}\rangle$ are not independent: as a consequence
of (\ref{n0-Gamma}) and (\ref{n0-C}) they satisfy
$n_{01;1}=\bar{n}_{01;2}$ and $n_{01;3}=\bar{n}_{01;3}$. Thus, if
$n_{01;3}\ne 0$  we can take $n_{01;3}=1$ and then only one
complex number, $n_{01;1}\equiv \rho_0$ determines

\b |n_{0,1}\rangle=(\rho_0, \bar{\rho}_0,1)^{T} \label{rho-0} \e
and therefore $|n_{0,j}\rangle$ and $|m_{0,j}\rangle$.

Let us denote $Q_0\equiv Q^*h_0$, where $h_0$ is the 'undressed'
potential. Then we have the following equation for $g$:

\b g_y+Qg-gQ_0+\lambda [g,J]=0, \label{Eq4g factor}\e satisfied
identically for any $z$, i.e. any $\lambda$. This equation is
satisfied for the construction (\ref{gfactor}) due to the fact
that

\b (\langle m_j|)_y -\langle m_j|(Q_0- \lambda_j J)&=&0, \\
(|n_j\rangle)_y + (Q_0- \mu_j J)|n_j\rangle &=& 0.\e

The identity (\ref{Eq4g factor}) for $\lambda \to \infty$ gives \b
Q=Q_0+[J, \alpha_1+\alpha_2 +\alpha_3]. \label{Eq4Q}\e
Multiplication of (\ref{Eq4Q}) by $Q^*$ followed by taking of a
trace gives

\b h(y,t)=h_0(y,t)+\frac{1}{2}\sum_{p,j=1}^{3}\hat{R}_{pj}\langle
m_{j}|[Q^*,J]|n_p \rangle. \label{h dressed}\e

The reality of (\ref{h dressed}) can be checked once again by
using the introduced properties of $R$, $|n_j\rangle$ and
$|m_j\rangle$.

Thus, to each discrete eigenvalue, which represents an 'action'
variable and is determined by only one real value, $\lambda_0$
(\ref{lambda-0}) one can put into correspondence a conjugated
'angle' variable that is given by the independent component of the
associated constant vector, $\rho_0$ (\ref{rho-0}). This accounts
for the scattering data related to the discrete spectrum. Since
$\lambda_0$ is real, in the $k$ - plane the corresponding discrete
spectrum value $k_0$ is also on the real line. It seems that $k_0$
is on the continuous spectrum, since $l_1$ and $l_4$ are on the
real line. However, the continuous spectrum for the actual
spectral problem (\ref{SP}) does not consist of the rays $l_{\nu}$
shown in Figure 1, but rather of the lines $\tilde{l}_{\nu}$ through
the origin, where $\arg(k)=\pi/6 + \nu \pi/3$. This is due to the
relation between the corresponding eigenfunctions (\ref{chi-nu})
involving nontrivial $k$-dependence in the factor $e^{\Lambda(k)
y}$.

To repeat the dressing procedure $N$ times we will need $N$ copies
of the same type scattering data, $\{\rho_{0,j}, \lambda_{0,j}\}$
$j=1, \ldots, N$. The result will be the $N$-soliton solution.

For example, the one-soliton solution can be obtained as follows.
One can start the dressing from the 'trivial' solution $u\equiv
0$, i.e. $h_0=1$. Then there is a global analytic solution of the
spectral problem, symbolically $e^{L_{\infty}(k)y+M_{\infty}(k)t}$, i.e.

\b \Psi(y,k;t)= U(k)e^{\Lambda
(k)y+A(k)t}\hat{U}(k). \nonumber \e

\n Since the last factor does not depend on $y$ and $t$ we can
take simply (for all sectors)

\b \chi_0(y,k;t)= U(k)e^{\Lambda (k)y+A(k)t}, \nonumber \e The entries of $U(k)$ are
\b  U_{1p}(k)&=&(\omega ^2-1) \lambda+(\omega-1)\Lambda_p+\omega^2, \nonumber \\
 U_{2p}(k)&=&(\omega -1) \lambda+(\omega^{2}-1)\Lambda_p+\omega,  \nonumber \\  U_{3p}(k)&=& 3(\Lambda_p^2+\lambda \Lambda_p + \lambda^2)-1. \nonumber \e

The potential for the one-soliton solution is given by (\ref{h
dressed}), then $f(y,t)$ ($t$ is viewed as an additional parameter
rather than a second independent variable) can be computed as a
solution of linear first order ODE (\ref{ODE}). Then $q(x,t)$ from
(\ref{q}) and therefore $u(x,t)$ can be obtained from
(\ref{m-parametric}) and (\ref{m}).

\section{Discussion}

In the presented analysis we formulated the inverse scattering
problem for the DP equation. We defined a set of scattering data
for the problem:

(i) on the continuous spectrum the coefficient
$[S_1^+(k)]_{12}$, $k \in l_1$;

(ii) on the discrete spectrum the set $\{ \rho_{0,j},
\lambda_{0,j}\}$, $j=1,\ldots, N$, where $N$ is the number of the
discrete eigenvalues.

The scattering data uniquely define the potential $h(y,t)$ and
therefore a general $N$-soliton Schwartz-class solution $u(x,t)$
of DP equation. As it is the case for all systems, integrable by
the Inverse Scattering Method, the mapping between the solution
and the scattering data allows the interpretation of a generalised
Fourier transform \cite{AKNS,GVY,GY,TV08}.

The peakon solutions (peaked solitons) appear in the limit $\kappa
\to 0$ \cite{L-S}, although such limit in the space of the
scattering data will require further considerations.

 The $N$-soliton solution of the Degasperis-Procesi equation is obtain in \cite{M1} by Hirota's method. E.g. the 1-soliton solution is
\b u(y,t)&=&\frac{\frac{8\kappa}{a_1}(a_1^2-1)(a_1^2-\frac{1}{2})}{\cosh \xi_1 +2a_1-\frac{1}{a_1}} \label{1-soliton} \\
x&=&y+\ln \left( \frac{\gamma_1+1+(\gamma_1-1)e^{\xi_1}}{\gamma_1-1+(\gamma_1+1)e^{\xi_1}}\right),\e
where the quantities $\xi_i$ are

\b \xi_i=\nu_i\left(y-\frac{3\kappa}{1-\nu_i^2}t -y_{i0} \right), \qquad i=1,2,\ldots, N, \label{dr} \e where $\nu_i$ and $y_{i0}$ are constants, representing the scattering data, \b a_i=\sqrt{\frac{1-\frac{1}{4}\nu_i^2}{1-\nu_i^2}} \nonumber \e as well as $\gamma_1=\sqrt{\frac{(2a_1-1)(a_1+1)}{(2a_1+1)(a_1-1)}}$ are also constants depending on the scattering data.

The quantities, related to the 1-soliton solution (\ref{1-soliton}) can also be computed:

\b f(y,t)&=&1+\frac{3 \nu_1^2e^{\xi_1}}{a_1(1-\nu_1^2)\left(1+\frac{2}{a_1}e^{\xi_1}+e^{2\xi_1}\right)} \nonumber \\
h(y,t)&=&\frac{\cosh 2 \xi_1 +\frac{4}{a_1}\cosh \xi_1 - \frac{3\nu_1^3}{a_1(1-\nu_1^2)}\sinh \xi_1}{\left(\cosh \xi_1 + \frac{2+\nu_1^2}{a_1(1-\nu_1^2)}\right)\left(\cosh \xi_1 + \frac{2}{a_1}\right)}.
\nonumber \e

One can establish that $\nu_i$ that determines the dispersion relation (\ref{dr}) is related to the spectral parameters,
\b \nu_i=\frac{1}{\sqrt{3}}\left(k_i+\frac{1}{k_i} \right), \nonumber \e where $k_i$ is a the real discrete eigenvalue, corresponding to the real $\lambda_{0,i}=\lambda(k_i)$.
The computation of the soliton solution by dressing method requires development of additional techniques addressing the technical difficulties arising in the computation.

There are interesting multidimensional versions of the DP
equation. They are in general nonintegrable, but admit singular
(peakon-type) solutions \cite{HS}.

\section*{Acknowledgements}

A.C. and R.I. acknowledge the support of the G. Gustafsson
Foundation for Research in Natural Sciences and Medicine (Sweden).
R.I. is also supported by Science Foundation Ireland, Grant
09/RFP/MTH2144. J.L. acknowledges support from a Marie Curie
Intra-European Fellowship. The authors are grateful to Prof. V.~S.~Gerdjikov and
Dr.~G.~Grahovski for many valuable discussions.

\bibliographystyle{plain}
\bibliography{is}

\begin{thebibliography}{99}
\small

\bibitem{AKNS}
M.~J.~Ablowitz, D.~J.~Kaup, A.~C.~Newell, and H.~Segur,
The inverse scattering transform --- Fourier analysis for nonlinear
problems, {\it Stud. Appl. Math.} {\bf 53} (1974), 249 --315.

\bibitem{AL}
B.~Alvarez-Samaniego and D.~Lannes, Large time existence for 3D water-waves and asymptotics,
{\it Invent. Math.} {\bf 171} (2008), 485--541.

\bibitem{AG06}
V.~Atanasov and  V.~Gerdjikov, On the multi-component nonlinear Schr\"odinger equation
with constant boundary conditions, in {\it Gravity, Astrophysics, and
Strings 05}, eds. P. P. Fiziev and M. D. Todorov, St. Kliment
Ohridski University Press, Sofia, 2006.

\bibitem{C-H}
R.~Camassa and D.~Holm, An integrable shallow water equation with
peaked solitons, {\it Phys. Rev. Lett.} {\bf 71} (1993), 1661--1664.

\bibitem{JC09}J. L.  Cie\'sli\'nski,  Algebraic construction of the Darboux matrix revisited, {\it J. Phys. A: Math. Theor.} {\bf 42} (2009) 404003.


\bibitem{C}
A.~Constantin, The trajectories of particles in Stokes waves,
{\it Invent. Math.} {\bf 166} (2006), 523--535.

\bibitem{CE}
A.~Constantin and J.~Escher, Wave breaking for nonlinear nonlocal shallow water equations,
{\it Acta Mathematica} {\bf 181} (1998), 229--243.

\bibitem{CE2}
A.~Constantin and J.~Escher, Particle trajectories in solitary water waves,
{\it Bull. Amer. Math. Soc.} {\bf 44} (2007), 423--431.

\bibitem{CGI}
A.~Constantin, V.~Gerdjikov and R.~Ivanov, Inverse scattering transform for the Camassa-Holm equation
{\em Inverse Problems} {\bf 22} (2006), 2197 -- 2207.

\bibitem{CL}
A.~Constantin and D.~Lannes, The hydrodynamical relevance of the Camassa-Holm
and Degasperis-Procesi equations, {\it Arch. Ration. Mech. Anal.} {\bf 192} (2009), 165--186.

\bibitem{D-P}
A.~Degasperis and M.~Procesi, Asymptotic integrability,
in {\it Symmetry and Perturbation Theory}, edited by A. Degasperis and G.
Gaeta, World Scientific (1999), pp. 23--37.

\bibitem{D-H-H}
A.~Degasperis, D.~Holm and A.~Hone, A new integrable equation with peakon
solutions, {\it Theor. Math. Phys.} {\bf 133} (2002), 1461--1472.

\bibitem{D-H-H2}
A.~Degasperis, D.~Holm and A.~Hone, Integrable and non-integrable
equations with peakons, in {\it Nonlinear Physics: Theory and
Experiment} (eds: M. Boiti et al.) World Scientific Publishing
2007, 37 -- 43.

\bibitem{ELY}
J.~Escher, Y.~Liu and Z.~Yin, Global weak solutions and blow-up structure
for the Degasperis-Procesi equation, {\it J. Funct. Anal.} {\bf 241} (2006), 457--485.

\bibitem{FaTa}
L.~D.~Faddeev and  L.~A.~Takhtadjan, {\it Hamiltonian methods in
the theory of solitons} (Springer Verlag, Berlin, 1987).

\bibitem{Ger}
V.~Gerdjikov, Algebraic and analytic aspects of soliton type
equations, \emph{Contemp. Math.} \textbf{301} (2002), 35--68.

\bibitem{VSG04}
V.~Gerdjikov, Basic aspects of soliton
theory, in {\it Sixth International Conference on Geometry,
Integrability and Quantization}, June 2004, Varna, Bulgaria,
ed: I. Mladenov and A. Hirshfeld, SOFTEX, Sofia 2005, pp.
78--125.

\bibitem{VSG86}
V.~Gerdjikov, Generalised Fourier transforms for the soliton
equations. Gauge covariant formulation., {\it Inverse Problems} {\bf 2} (1986), 51--74.

\bibitem{VSG1}
V.~S.~Gerdjikov, The Zakharov-Shabat dressing method and the
representation theory of the semisimple Lie algebras, {\it Phys. Lett.
A} {\bf 126} (1987), 184--188.


\bibitem{GG1}
V.~S.~Gerdjikov, G.~G.~Grahovski and N.~A.~Kostov,
Reductions of $N $-wave interactions related to low rank Lie
algebras I: $ Z_2$ -- reductions, {\it J. Phys. A} {\bf 34} (2001),  9425--9461.

\bibitem{GG2}
V.~S.~Gerdjikov and G.~G.~Grahovski, On N-wave type and
NLS type systems and their gauge equivalent: generating operators
and the gauge group action, {\it Proc. NAS Ukraine} {\bf
50} (2004), 388--395.


\bibitem{GGK05a}
V.~S.~Gerdjikov, G.~G.~Grahovski and N.~A.~Kostov, On the
multi-component NLS type equations on symmetric spaces and their
reductions, {\it Theor. Math. Phys.} {\bf 144} (2005), 1147--1156.

\bibitem{GK83}
V.~Gerdjikov and P.~P.~Kulish, The multicomponent nonlinear
Schr\"odinger equation in the case of nonzero boundary conditions,
{\it Zap. Nauchn. Sem. Leningrad. Otdel. Mat. Inst. Steklov} \textbf{131} (1983), 34--46 (in Russian);  {J.
Math. Sci.} \textbf{30} (1985), 2261--2269 (in
English).


\bibitem{GKV}
V.~Gerdjikov, N.~Kostov and T.~Valchev, Soliton equations with deep
reductions. Generalized Fourier transforms, in {\it  Topics in
contemporary differential geometry, complex analysis and
mathematical physics},  pp. 85--96, World Sci. Publ., Hackensack, NJ,
2007.

\bibitem{GVY}
V.~Gerdjikov, G.~Vilasi and A.~Yanovski, {\it Integrable
Hamiltonian Hierarchies. Spectral and Geometric Methods}, Lecture
Notes in Physics 748, Springer, Berlin - Heidelberg, 2008.

\bibitem{GY}
V.~Gerdjikov and A.~B.~Yanovski, Completeness of the
eigenfunctions for the Caudrey-Beals-Coifman system, \emph{J.
Math. Phys.}  \textbf{35} (1994), 3687--3725.


\bibitem{G1}
G.~G.~Grahovski and M.~Condon, On the Caudrey-Beals-Coifman System and the Gauge Group Action,
{\it J. Nonlin. Math. Phys.} {\bf 15} (2008), Suppl. 3, 197--208.


\bibitem{G2}
G.~G.~Grahovski, On the reductions and scattering
data for the CBC system, in {\it Geometry, Integrability and
Quantization III}, Eds:  I. Mladenov and G.  Naber, Coral Press,
Sofia, 2002, pp. 262--277.

\bibitem{G3}
G.~G.~Grahovski, On the reductions and scattering
data for the generalized Zakharov--Shabat systems, in {\it Nonlinear
Physics: Theory and Experiment. II}, Eds:  M.~J.~Ablowitz, M.~Boiti,
F.~Pempinelli and B.~Prinari, World Scientific, Singapore, 2003, pp.
71--78.

\bibitem{G4}
G.~G.~Grahovski, V.~S.~Gerdjikov, N.~A.~Kostov, V.~A.~Atanasov, New integrable
multi-component NLS type equations on
symmetric spaces: $\mathbb{Z}_4 $ and $\mathbb{Z}_6 $ reductions,
in {\it Geometry, Integrability and Quantization VII}, Eds: I.~Mladenov and M.~De Leon, Softex, Sofia (2006), 154--175.


\bibitem{H}
D.~Henry, Persistence properties for a family of nonlinear partial differential equations,
{\it Nonlinear Anal.} {\bf 70} (2009), 1565--1573.

\bibitem{HS} D. Holm, M. Staley, Wave structure and nonlinear balances in a family of evolutionary
PDEs, {\it SIAM J. Appl. Dyn. Syst.}  {\bf 2} (2003), 323--380.

\bibitem{HW}
A.~N.~W.~Hone and J.~P.~Wang, Prolongation algebras and
Hamiltonian operators for peakon equations, \emph{Inverse
Problems} {\bf 19} (2003), 129--145.

\bibitem{Ho}
L.~H\"ormander, {\it The analysis of linear partial differential operators. I. Distribution theory and Fourier analysis},
Springer-Verlag, Berlin, 2003.



\bibitem{I}
R.~I.~Ivanov, On the integrability of a class of nonlinear dispersive wave equations,
{\it J. Nonlinear Math. Phys.} {\bf 12} (2005), 462--468.

\bibitem{I2}
R.~I.~Ivanov, Water waves and integrability, {\it Philos. Trans. Roy. Soc. London A}
{\bf 365} (2007), 2267--2280.


\bibitem{I3}
R.~I.~Ivanov, On the dressing method for the generalised
Zakharov-Shabat system, {\it Nucl. Phys. B} {\bf 694} (2004),
509--524.

\bibitem{J0}
R.~S.~Johnson, {\it A modern introduction to the mathematical theory of water waves},
Cambridge University Press, Cambridge, 1997.

\bibitem{J}
R.~S.~Johnson, Camassa-Holm, Korteweg-de Vries and related models for water waves,
{\it J. Fluid Mech.} {\bf 455} (2002), 63--82.

\bibitem{Ka}
D.~J.~Kaup, On the inverse scattering problem for cubic eigenvalue problems of the class $\psi \sb{xxx}+6Q\psi \sb{x}+6R\psi =\lambda \psi $, {\it Stud. Appl. Math.} {\bf 62} (1980), 189--216.

\bibitem{Ka2}
D.~J.~Kaup, The legacy of the IST, in {\it The legacy of the inverse scattering transform in applied mathematics}
(South Hadley, MA, 2001), pp. 1--14, Contemp. Math., 301, Amer. Math. Soc., Providence, RI, 2002.

\bibitem{L}
J.~Lenells, Traveling wave solutions of the Degasperis-Procesi equation,
{\it J. Math. Anal. Appl.} {\bf 306} (2005), 72--82.

\bibitem{LY}
Y.~Liu and Z.~Yin, Global existence and blow-up phenomena for the
Degasperis-Procesi equation, {\it Comm. Math. Phys.} {\bf 267} (2006),
801--820.

\bibitem{L-S}
H.~Lundmark and J.~Szmigielski, Multi-peakon solutions of the
Degasperis-Procesi equation, {\it Inverse Problems} {\bf 19}
(2003), 1241--1245.

\bibitem{M1}
Y.~Matsuno, The $N$-soliton solution of the Degasperis-Procesi equation, {\it
Inverse Problems} {\bf 21} (2005), 2085--2101.

\bibitem{M2}
Y.~Matsuno, Multisoliton solutions of the Degasperis-Procesi equation and their peakon limit, {\it Inverse Problems} {\bf 21} (2005), 1553--1570.

\bibitem{AV}
A.~V.~Mikhailov, The reduction problem and the inverse
scattering method, {\it Physica} \textbf{3D} (1981), 73.

\bibitem{ZMNP}
S.~P.~Novikov, S.~V.~Manakov, L.~P.~Pitaevskii, and V.~E.~Zakharov,
{\it Theory of solitons: the inverse scattering method}, New
York: Plenum, 1984.

\bibitem{Sh75}
A.~B.~Shabat, The inverse scattering problem for a system of
differential equations, {\it Funkcional. Anal. i Prilozen.} {\bf 9} (1975), 75--78 (in Russian).

\bibitem{Sh79} A.~B.~Shabat, An inverse scattering problem,
{\it Diff. Uravneniya} {\bf 15} (1979), 1824--1834 (in Russian).

\bibitem{To}
J.~F.~Toland, Stokes waves, {\it Topol. Methods Nonlinear Anal.} {\bf 7} (1996), 1--48.


\bibitem{TV08} T.~Valchev, On the Kaup-Kupershmidt equation. Completeness
relations for the squared solutions, in {\it Ninth International Conference on Geometry, Integrability
and Quantization, June 8-13 2007, Varna, Bulgaria}, ed: I.
Mladenov and M. de Leon, SOFTEX, Sofia 2008, pp. 1--12.


\bibitem{ZaSh1}
V.~Zakharov and A.~Shabat, A scheme for integrating the nonlinear equations of mathematical
 physics by the method of the inverse scattering problem - I, {\it Func. Anal. Appl.} {\bf 8} (1974), 226--235.

\bibitem{ZS1}
V.~Zakharov and A.~Shabat, Integration of nonlinear
equations of mathematical physics by the method of inverse
scattering - II, {\it Func. Anal. Appl.} \textbf{13} (1979), 166--174 (English translation).



\end{thebibliography}

\end{document}